# STOCHASTIC MODELS OF MISINFORMATION DISTRIBUTION IN ONLINE SOCIAL NETWORKS

Konstantin Abramov, Yuri Monakhov

*Department of Informatics and Information Security, Vladimir State University, Vladimir 600000 Russian Federation*

***This report contains results of an experimental study of the distribution of misinformation in online social networks (OSNs). We consider the classification of the topologies of OSNs and analyze the parameters identified in order to relate the topology of a real network with one of the classes. We propose an algorithm for conducting a search for the percolation cluster in the social graph.***

Results of current research show the intensive growth in OSN users' activity [1]. This phenomenon is observed throughout the entire Internet. Currently there are many factors enabling massive waves of misinformation spread through different social media. Finding ways to mitigate these phenomena is an extremely actual and very complicated problem. One of the aspects that we need to take into account is building adequate models and prediction methods for the misinformation spread. The easiest and most researched way to model (and predict) these processes, in our opinion, is to use the classic epidemiological models developed in 19th century to study how infectious diseases spread among the population. These models generally are based on systems of differential equations. Such models though are rather primitive and they don't take into the account some of the vital characteristics of the misinformation process[2]. Therefore the development of the new models is relevant and necessary to predict the spread of misinformation and to mitigate it.

Researchers present different classifications of the network topologies and their characteristic properties. According to one of these classifications[3], there are three different types of small world networks: wide scale, single scale and scale free (SF) networks. The first type is characterized by the node degree distribution function, which is a mix of power-law and exponential distributions. The primary characteristic of the second type is an absence (or very low percent) of the nodes with high degree, i.e. Watts-Strogatz networks. Finally, the scale free networks are strictly described by the power-law node degree distribution, meaning the probability of the nodes with high degree of connectivity is relatively high. The most known example of the SF networks is the Barabasi-Albert (BA) network.

The table (Tab. 1) describes features one can use to relate the OSN topology to one of the types[4]. These features are:
- *the number of connections in the network with N nodes* – the number of all possible connections between N nodes;
- *the average path length l* between any pair of vertices (nodes);
- *the node connectivity degree distribution function* shows the probability of the node having the degree of *k;*
- *the clustering coefficient* that is defined as the probability of a connection between two nodes having a common neighbor. This probability (for an i-th neighboring node) is calculated as

$$C_i = \frac{E_i}{\frac{1}{2}k_i(k_i-1)},$$

where $k_i$ is the i-th neighboring node degree and $E_i$ is the total number of connections between k nodes adjacent to the i-th node. The *total* (or *average*) *clustering coefficient* for a network with N nodes will be

$$C = \frac{1}{N}\sum_i C_i.$$

**Tab.1**

**Basic topological features**

| Feature | Watts-Strogatz small world network | Barabasi-Albert scale free network | Random networks (i.e. Erdos-Renyi graph) |
|---|---|---|---|
| Number of connections | $\dfrac{k \cdot N}{2}$ | $m \cdot (N-1)$ | $\dfrac{\langle k \rangle \cdot N}{2}$ |
| Average path length | $\sim \dfrac{\ln N}{\ln k}$ | $m=1: l \sim \ln N;$ $m \geq 2: l_{BA}^{\alpha>3} \approx \ln N;$ $l_{BA}^{\alpha=3} \approx \ln N / \ln \ln N;$ $l_{BA}^{2<\alpha<3} \approx \ln \ln N.$ | $\sim \dfrac{\ln N}{\ln \langle k \rangle}$ |
| Clustering coefficient | $C_{p \to 1} \sim k/N,$ $C_{p>0} \gg C_{p \to 1}$ | $C \sim 5\dfrac{\langle k \rangle}{N}$ | $C \sim \dfrac{k}{N}$ |
| Node degree distribution | Poisson's law: $p(k) \sim e^{-pN}\dfrac{(pN)^k}{k!}$ | Power law: $p(k) \sim k^{-\alpha}$ | Poisson's law: $p(k) \sim e^{-pN}\dfrac{(pN)^k}{k!}$ |

Gubanov, Novikov and Chkhartishvili in their book [5] provide an extensive review of the approaches towards social network modeling. In their research they classify netwoks not by their structure but rather by the types of social influence. On the other hand their statement concerning the necessity of having at least two features to identify the network type is well applicable in our case. We select the node degree distribution function and the clustering coefficient for these two features.

During the research we had extracted the topological data of the *Facebook* OSN, the size of the obtained sample equals 6927 nodes. The following research was conducted on this subgraph.

The authors have conducted a comparison between the node degree distribution acquired from the Facebook OSN subgraph and the degree distributions of small world, scale free and Erdos-Renyi networks. For Erdos-Renyi and small world networks one can approximate the degree distribution with Poisson's law, and for scale free networks the distribution of the degrees is approximated by power law. In our social subraph maximum degree equals 507, minimum degree equals 1. The histogram of the node degree distribution of the sample is shown in Fig. 1

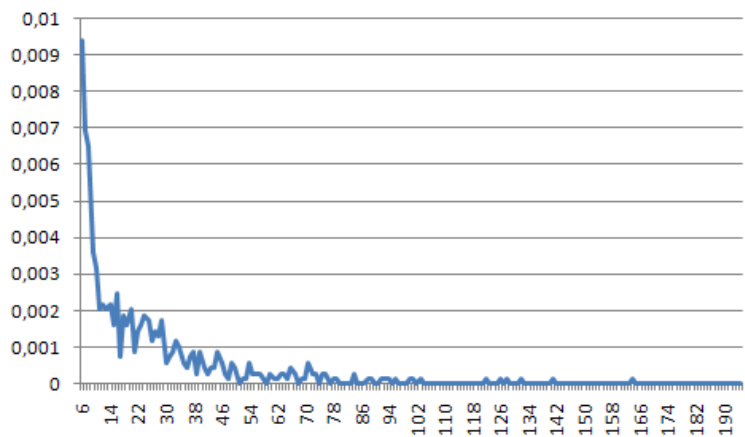

Fig. 1. Node degree distribution of the sample

It is well known that the characteristic feature of scale free networks is the presence of a few nodes with very high degrees of connectivity and a mass of nodes with small connectivity degrees. We can presume that the obtained sample is scale free. To test this hypothesis the probabilities of a node having different degrees were calculated. The results show that the acquired dataset can be (with sufficient accuracy, $\Delta=0,01$) approximated by a power function. Therefore the degree distribution function for our social subgraph is that of a scale free network.

The value of a clustering coefficient calculated for this subgraph complies with the criterion shown in Tab. 1, which is typical for the scale free networks. We therefore can conclude that our Facebook OSN sample has the scale free topology[6].

During this research we carried out the simulation of the misinformation process in OSNs. The simulation was conducted using proprietary softwrae developed by authors. The most commonly used models of social influence are the classic epidemic models, such as SI, SIR, SIS, PSIDR, AAWP, SIM. Every model has its specific features and no model is universally applicable. In this research we use SIR model as the source.

The most significant result we achieve from the simulation is that in most cases the cluster that consists of R-nodes (nodes "immune" to misinformation) significantly reduces the speed of spread by isolating the "infected" nodes or spreaders of misinformation from the nodes "susceptible" to misinformation. We use the term "cluster" to describe the linked subgraph constructed from the nodes with high clustering coefficients.

**REFERENCES**

[1] Industry report of the Federal Agency for Press and Mass Communications "Internet in Russia: Status, Trends and Prospects", August 2010. [Russian: Отраслевой доклад Федерального агентства по печати и массовым коммуникациям «Интернет в России: состояние, тенденции и перспективы развития», август 2010г.]

[2] Konstantin Abramov, Yuri Monakhov. Modeling of unwanted information in social media [text] / Proceedings of XXX All-Russian Conference on problems of efficiency and safety of complex engineering and information systems. Part IV, section number 6. - Serpukhov VI RV. - 2011. - 376 p.; - P. 178-182. - ISBN 978-5-91954-029-8 [Russian: Абрамов К.Г., Монахов Ю.М. Моделирование распространения нежелательной информации в социальных медиа [Текст] // Труды XXX Всероссийской научно-технической конференции. Проблемы эффективности и безопасности функционирования сложных технических и информационных систем. Часть IV, секция №6. - Серпуховский ВИ РВ. - 2011. - 376 с.; - С. 178-182. - ISBN 978-5-91954-029-8]

[3] Leveille J. Epidemic Spreading in Technological Networks // M. Sc. Thesis, HP Labs Bristol, 2002 – 100 pp.

[4] Zharinov IV, Krylov VV. Construction of graphs with minimum average path length / / Bulletin of Izhevsk State Technical University - № 4, 2008. -P. 164-169. [Russian: Жаринов И.В., Крылов В.В. Конструирование графов с минимальной средней длиной пути // Вестник Ижевского государственного технического университета -№4, 2008. –С. 164-169. ]

[5]. Gubanov DA, Novikov DA, Chkhartishvili AG. Models of influence in social networks, overview of classes of modern models of social networks, 2004. [Russian: Губанов Д.А., Новиков Д.А., Чхартишвили А.Г. «Модели влияния в социальных сетях», обзор классов современных моделей социальных сетей, 2004.]

[6]. Yuri Monakhov, Konstantin Abramov, Alexandra Nikitashenko. On the question of improving the models of misinformation spread in online social networks // Information Systems and Technology IST-2011: Proceedings of the XVII International Scientific and Technical Conference, Nizhny Novgorod, 2011 [Russian: Абрамов К.Г., Монахов Ю.М., Никиташенко А.В. К вопросу об уточнении моделей распространения нежелательной информации в социальных сетях Интернета / Информационные системы и технологии ИСТ-2011: материалы XVII международной научно-технической конференции - Н. Новгород: - 2011. - ISBN 978-5-9902087-2-8]